\documentclass[runningheads]{llncs}
\usepackage{threeparttable}
\usepackage{tabu}                     
\usepackage{multirow}                 
\usepackage{multicol}                 
\usepackage{float}                    
\usepackage{makecell}                 
\usepackage{booktabs}                 
\usepackage{graphicx}
\usepackage{subfigure}
\usepackage{amsmath}
\usepackage{amsfonts}
\usepackage{amssymb}
\usepackage[ruled,linesnumbered]{algorithm2e}
\usepackage[T1]{fontenc}
\usepackage{balance}

\begin{document}
\title{Beyond the Trigger: Learning Collaborative Context for Generalizable Trigger-Induced Recommendation}

\titlerunning{Collaborative Contrastive Network}

\author{Chen Gao\inst{1}\orcidID{0009-0005-6417-1428} \and
Zixin Zhao\inst{1}\orcidID{0009-0006-8638-7534} \and
Lv Shao\inst{1}\orcidID{0009-0008-5389-6476} \and
Tong Liu\inst{1}\orcidID{0000-0003-2425-0357} }

\authorrunning{C. Gao et al.}
\institute{Alibaba Group, Hangzhou, China
\email{\{gaochen.gao,foriyte.zzx,shaolv.sl,yingmu\}@taobao.com}}

\maketitle
\begin{abstract}
In e-commerce, Trigger-Induced Recommendation (TIR), recommending items after a user clicks a trigger, is an important task. However, modern platforms rely on a continuous stream of diverse and short-lived promotional scenarios (e.g., for Black Friday), creating a significant challenge. Existing methods are less effective here: they either fall into a trigger-dependency trap, recommending overly similar items, or a data-hungry trap, requiring long-term stable data for intent modeling that these ephemeral scenarios cannot provide.

To address these limitations, we propose the Collaborative Contrastive Network (CCN), a general and robust framework that approaches the problem from a different perspective. Instead of modeling ambiguous entry intent, CCN learns a user's context-specific preferences by treating the user-trigger pair as a unique condition. It achieves this via a novel contrastive learning scheme, using the collaborative feedback of co-click/co-non-click as a positive signal and mono-click as a negative signal to structure the item representation latent space. To prove its real-world generality, CCN is trained on a heterogeneous dataset spanning over a dozen different scenarios from an entire year, and the online A/B test is conducted in a completely new, unseen scenario on Taobao, where CCN boosts CTR by 12.3\% and order volume by 12.7\%, demonstrating its effectiveness and generalization.
% \keywords{CTR prediction \and Recommender Systems \and Generative Model \and Temporal Modeling}
\end{abstract}
\section{Introduction}

In modern e-commerce platforms, specialized recommendation scenarios, such as New Product Channels or promotional mini-apps for sales events like Black Friday, have become pivotal for user engagement and conversion. A common user journey begins with clicking a trigger item displayed at an entrance, which then leads to a dedicated page of related items. This paradigm, which we term Trigger-Induced Recommendation (TIR), presents a unique challenge: the trigger item not only reflects a user's immediate interest but also establishes a strong contextual bias for all subsequent interactions within that session. The core task is to accurately predict the Click-Through Rate (CTR) for these subsequent items, conditioned on the specific trigger that initiated the session.

\begin{figure}[tbp]
  \centering
  \includegraphics[width=0.95\linewidth]{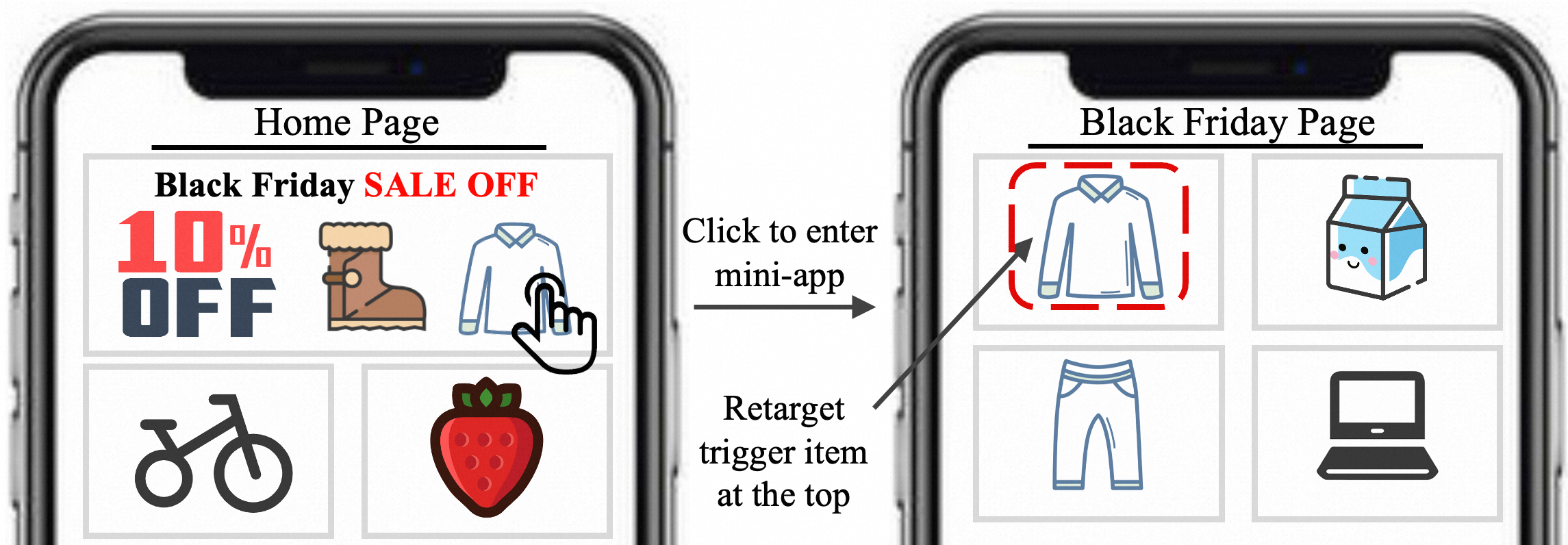}
  \caption{Trigger-Induced Recommendation in Mini-Apps}
  \label{fig:shouji_new}
\end{figure}

Conventional CTR models Wide\&Deep~\cite{WideDeep}, DIN~\cite{DIN}, DIEN~\cite{DIEN}, DSIN~\cite{DSIN}, SIM~\cite{SIM}, Gift~\cite{cao2022gift} are not directly suited for the TIR task. By treating the trigger as just another feature, they fail to capture its special role as a context-setter that dynamically reshapes a user's interest, leading to suboptimal performance.

To address this, trigger-aware models have been proposed, but they fall into two traps, leaving critical weakness for scenarios that are both dynamic and short-lived.
(i) The Trigger-Dependency Trap: One line of work DIHN~\cite{DIHN} operates under the strong assumption that there is a necessary causal relationship between the user's entry and the trigger item. This causes the model to over-rely on the trigger, primarily recommending items highly similar to it. This approach may not fully capture the user's true intent when their interest is actually in the broader theme of the mini-app (e.g., Black Friday deals) rather than the specific trigger.
(ii) The Data-Hungry Intent-Modeling Trap: Another line of work DIAN~\cite{DIAN} attempts to explicitly disentangle a user's entry intent into a binary classification between being trigger-driven or app-driven. While conceptually appealing, this requires the mini-app to exist for a long time to accumulate sufficient data for stable intent modeling. This renders it impractical and non-robust for the most valuable and dynamic mini-apps, such as those created for ephemeral sales events like the Double 11 Shopping Carnival.

In this paper, we argue that modeling a user's binary entry intention can be a challenging premise. Instead of guessing why a user entered, we propose to model what they prefer once they are inside. Our core insight is that the user's clicks and non-clicks on the items displayed together on a single page provide a direct self-supervisory signal that directly reflects their preferences under the specific context of the trigger.

To leverage this signal, we introduce the Collaborative Contrastive Network (CCN), a general and robust framework for TIR. The central mechanism of CCN is to treat the user-trigger pair as a composite key that defines a unique contextual space for user interest. Within this specific context, CCN implements our insight through a novel collaborative contrastive learning paradigm. We leverage the collaborative relationship of co-click/co-non-click as a positive signal, pulling these items together in the latent space. Conversely, the non-collaborative relationship of mono-click (one item clicked, one not) serves as a negative signal, pushing them apart. This process forces the model to learn two distinct item clusters representing the user's context-specific interests and disinterests, directly in the latent space shaped by the trigger.

This paradigm directly addresses the challenges of prior work. It is general enough to work without long-term interaction data and robust enough to avoid being misled by the trigger item. During training, we further refine this process by using an importance sampling strategy to stabilize training and by re-weighting the contrastive loss to handle the natural sparsity of click data.

To rigorously validate the real-world robustness and generality of CCN, we design a demanding experimental setup. Our training dataset is constructed from a highly heterogeneous collection of over a dozen distinct mini-app scenarios from Taobao within a single year. These scenarios vary significantly in their marketing intents (e.g., new arrivals vs. clearance sales), target user group, and overall data distributions. Critically, to prove its generalization capability, both offline experiments and the online A/B test are conducted in a completely new, previously unseen scenario.

Our main contributions are summarized as follows:
\begin{itemize}
\item We identify two fundamental traps in existing TIR models: trigger-dependency and the reliance on long-term data. We propose a new perspective that bypasses ambiguous intent modeling by focusing on in-session collaborative signals.

\item We propose the Collaborative Contrastive Network (CCN), a general and robust framework that models user's trigger-conditioned interests by treating the user-trigger pair as a unique context and contrasting collaborative versus non-collaborative behavioral feedback.

\item Extensive experiments are conducted on a heterogeneous dataset. Our model demonstrates exceptional generalization, boosting CTR by 12.3\% and order volume by 12.7\% in a previously unseen scenario during online A/B Test over the production baseline.
\end{itemize}

\section{Proposed Method}

\begin{figure*}[tbp]
\centering
\includegraphics[width=0.97\linewidth]{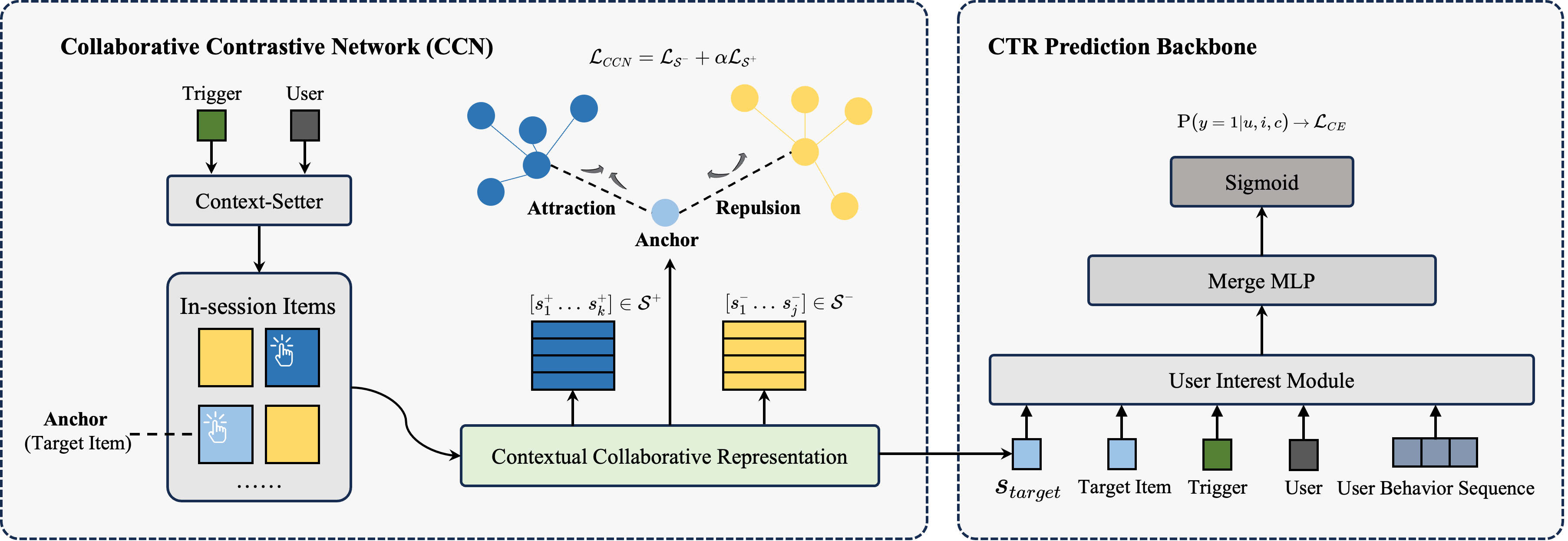} 
\caption{The overall architecture of our model. It consists of a CTR Prediction Backbone and our core innovation, the Collaborative Contrastive Network (CCN). CCN learns a trigger-conditioned collaborative representation which is then fed as a critical feature into the backbone to enhance the final prediction.}
\label{CCN}
\end{figure*}

\subsection{Overall Architecture}

As illustrated in Figure~\ref{CCN}, our model is composed of two main components:
\begin{itemize}
    \item A CTR Prediction Backbone: This is a standard deep learning architecture responsible for processing features and making the final CTR prediction.
    \item The Collaborative Contrastive Network (CCN): This is the core of our proposed method. Its primary goal is to learn a context-aware representation for each item based on in-session collaborative signals.
\end{itemize}

The representation learned by the CCN module is fed as an additional, highly informative feature into the CTR Prediction Backbone. This allows the main prediction task to directly benefit from the fine-grained, context-specific user preferences captured by our collaborative contrastive learning paradigm. The entire model is trained end-to-end through a joint optimization of the CTR prediction loss and the contrastive loss from CCN.

\subsection{Collaborative Contrastive Network (CCN)}

The CCN module is designed to address the core challenges of TIR by learning from in-session collaborative signals. It consists of two key stages: generating contextual collaborative representations and optimizing them with a tailored contrastive objective.

\subsubsection{Contextual Collaborative Representation.}

Instead of using a static item embedding, we aim to generate a dynamic representation for each item that is conditioned on the specific context, i.e., the user and the trigger item. We term this representation the Collaborative Representation, denoted by $s$.

The inputs to this module are the embedding of an anchor item $E$ (either the target item or a context item), the trigger item embedding $E_{trigger}$, and the user profile embedding $E_{user}$. To capture the complex interactions, we formalize the collaborative representation $s$ as:

\begin{equation}
\label{collaborativedegree}
s = \mathrm{MLP}(E_{user} \oplus (E \odot E_{trigger})),
\end{equation}

where $\oplus$ denotes concatenation and $\odot$ denotes the Hadamard product. For efficiency, item embeddings $E$ are composed of streamlined features such as item id, category id and seller id.

\subsubsection{Collaborative Contrastive Objective.}
The core idea of CCN is to structure the latent space of these collaborative representations based on in-session user feedback. For a given anchor item $E$, we partition the other items on the same page into two sets:
\begin{itemize}
    \item Positive Set $\mathcal{S}^{+}$: Items that have the same click label as the anchor (co-clicked or co-unclicked). They represent collaborative feedback.
    \item Negative Set $\mathcal{S}^{-}$: Items that have a different click label from the anchor (mono-clicked). They represent non-collaborative feedback.
\end{itemize}

We then design a contrastive objective loss function $\mathcal{L}_{CCN}$ consisting of a repulsion loss $\mathcal{L}_{\mathcal{S}^{-}}$ and an attraction loss $\mathcal{L}_{\mathcal{S}^{+}}$. Note that a page here refers to the entire channel where all items with real exposure belong to a single context regardless of scrolling.

\subsubsection{Repulsion Loss.} To push the anchor away from items in the negative set $\mathcal{S}^{-}$, we define the repulsion loss. It is a variant of InfoNCE~\cite{infoNCE}, enhanced with an importance sampling mechanism to introduce mutual information:
\begin{equation}
\mathcal{L}_{\mathcal{S}^{-}} = - \mathbb{E}_{i \in \mathcal{D}} \left[ \mathrm{log}\frac{e^{s_{i}/\tau}}{e^{s_{i}/\tau} + \sum_{s'_{j} \in \mathcal{S}_{i}^{-}} \omega_{j}^{-} e^{s'_{j}/\tau}} \right],
\end{equation}

where $\tau$ is a temperature hyperparameter and $\omega_{j}^{-}$ is the importance sampling weight, which assigns refined weights to outliers with high collaborative degrees.

\subsubsection{Attraction Loss.} To pull the anchor closer to items in the positive set $\mathcal{S}^{+}$, we use an attraction loss based on cosine similarity:
\begin{equation}
\mathcal{L}_{\mathcal{S}^{+}} = \mathbb{E}_{i \in \mathcal{D}} \left[ 1 - \mathrm{cos}(s_{i}, \sum_{s'_{k} \in \mathcal{S}_{i}^{+}} \omega_{k}^{+} s'_{k}) \right],
\end{equation}
where $\omega_{k}^{+}$ is the importance sampling weight. We use cosine similarity considering it is a standard metric for attraction and is bounded, contributing to training stability.

For both positive and negative sets, the importance sampling weight $\omega$ is calculated within their respective sets:
\begin{equation}
\omega = \frac{e^{-s/\tau}}{\sum_{s_{i}\in{\mathcal{S}}}e^{-s_{i}/\tau }}.
\end{equation}

\subsection{CTR Prediction and Joint Optimization}
\subsubsection{Input Representation.}
The CTR Prediction Backbone takes categorical, numerical and sequence features as input and respectively projects them into dense embeddings for tensor calculation. 

\subsubsection{User Interest Modeling.}
To capture general user interests, we process the user's short-term and long-term behavior sequences. Following standard practices in DIN~\cite{DIN} and SIM~\cite{SIM}, we employ extra multi-head target attention mechanism and search-based sequence modeling especially on trigger item to extract user's immediate interest representations, denoted as $\textbf{\textit{H}}_{uim}$. Since this is not the core contribution of our work, we omit the detailed formulations.

\subsubsection{Final Prediction and Joint Training.}
The final representation for prediction is formed by concatenating all relevant embeddings, including the collaborative representation $s$ learned from our CCN module:

\begin{equation}
\label{final_predict_revised}
\begin{split}
\hat{y} = \mathrm{Sigmoid}\big( & \mathrm{MLP} (E_{user} \oplus E_{target} \oplus  E_{trigger} \oplus  \\
                               & \qquad \textbf{\textit{H}}_{uim} \oplus  s_{target})  \big).
\end{split}
\end{equation}

The main CTR prediction task is optimized using the binary cross-entropy loss $\mathcal{L}_{CE}$. The entire model is trained end-to-end with a joint loss function, combining the CTR loss and our proposed collaborative contrastive loss:
\begin{equation}
\mathcal{L} = \mathcal{L}_{CE} + \alpha \mathcal{L}_{\mathcal{S}^{-}} + \beta \mathcal{L}_{\mathcal{S}^{+}},
\end{equation}

\begin{equation}
\mathcal{L} = \mathcal{L}_{CE} + \lambda \mathcal{L}_{CCN},
\end{equation}
where $\mathcal{L}_{CCN} = \mathcal{L}_{\mathcal{S}^{-}} + \alpha \mathcal{L}_{\mathcal{S}^{+}}$ and $\lambda$, $\alpha$ are hyperparameters. The weight $\alpha$ is used to balance the attraction and repulsion tasks, which can be empirically set or determined by the natural ratio of positive to negative pairs.

\section{Inference and Deployment}
Our model is designed for efficient, large-scale industrial deployment. During online inference, only the CTR Prediction Backbone is executed, using the learned embeddings and the collaborative representation $s$ for the target item to be scored. The computationally intensive collaborative contrastive learning branch, including the processing of context items, is defined within a training-only operator. This ensures that the online serving latency is not affected by our proposed auxiliary training task, making CCN a practical and scalable solution for real-world systems.

\section{Experiments}

\subsection{Experimental Setup}

\subsubsection{Datasets.}
To evaluate our model's generalization capability in dynamic, short-lived scenarios, we construct a demanding dataset from Taobao's industrial logs.
\begin{itemize}
    \item Training Set: A large-scale, heterogeneous dataset comprising over a dozen different promotional scenarios spanning an entire year. This forces the model to learn generalizable patterns rather than memorizing features of a single scenario.
    \item Test Set: Data from a completely new, previously unseen scenario: the Taobao Double 11 Shopping Carnival. This setup provides the rigorous test for generalization, as the model has no prior exposure to this specific event's data distribution.
\end{itemize}
Detailed statistics of the dataset are presented in Table \ref{tab:dataset}.

\begin{table}[ht]
  \centering
  % \vspace{0.2cm}
  \caption{ Statistics of the industrial dataset.}
\normalsize
    \begin{tabular}{llll}
    \toprule
    Dataset & \#Users & \#Items & \#Instances\\
\hline
     Train & 61.6M & 19.1M & 2.33B \\
     Test & 56.2M & 10M & 536.8M \\
     \bottomrule
    
    \end{tabular}
  \label{tab:dataset}
\end{table}

\subsubsection{Baselines and Implementation Details.}

We compare CCN against two types of models: (1) widely-used general CTR models (Wide\&Deep, DIN, SIM) that are unaware of the trigger's special role, and (2) recent popular trigger-aware models (DIHN, DIAN). SIM serves as the online baseline. For a fair comparison, all models are implemented on a unified framework with identical features.

\subsection{Offline Performance}
We evaluate the models using the AUC~\cite{DIN} metric. As shown in Table \ref{tab:auc}, CCN outperforms all baselines. Notably, existing trigger-aware models like DIHN and DIAN show degraded performance on our test set. This reveals that these models are not robust enough for new, short-lived scenarios. For instance, the intention estimation auxiliary task in DIAN yields an AUC of 0.60 in our setting, compared to the 0.94 observed in its original evaluation, highlighting limitations in its cross-domain generalizability. In contrast, CCN avoids explicit intent modeling, achieving more consistent performance and enhanced robustness.

\begin{table}[t]
  \centering
  \caption{Offline performance comparison (AUC) and ablation study on our industrial test set. The results are averaged over five independent runs. $\Delta$AUC shows the improvement over the strong baseline (SIM).}
  \label{tab:auc}
  \small
  \setlength{\tabcolsep}{4pt}
  \begin{tabular}{l l c}
    \toprule
    \textbf{Category} & \textbf{Model} & \textbf{AUC} (\textbf{$\Delta$AUC}) \\
    \midrule
    \multicolumn{3}{l}{\textit{(A) General CTR Models}} \\
    & Wide\&Deep & 0.6872 ( - ) \\
    & DIN            & 0.7117 ( - ) \\
    & SIM (Baseline)  & 0.7276 ( - ) \\
    \midrule
    \multicolumn{3}{l}{\textit{(B) Trigger-Aware Models}} \\
    & DIHN           & 0.7259 ( -0.17pt ) \\
    & DIAN          & 0.7200 ( -0.76pt ) \\
    \midrule
    \multicolumn{3}{l}{\textit{(C) Ablation Study of Our Method}} \\
    & 1. Baseline + Trigger   & 0.7281 ( +0.05pt ) \\
    & 2. Baseline + Trigger + UIM & 0.7293 ( +0.17pt ) \\
    & 3. \textbf{CCN (Ours)}  & \textbf{0.7326} ( \textbf{+0.50pt} ) \\
    & \quad - w/o Attraction Loss ($\mathcal{L}_{\mathcal{S}^{+}}$) & 0.7312 ( +0.36pt ) \\
    & \quad - w/o Repulsion Loss ($\mathcal{L}_{\mathcal{S}^{-}}$)  & 0.7309 ( +0.33pt ) \\
    & \quad - w/o UIM       & 0.7315 ( +0.39pt ) \\
    \bottomrule
  \end{tabular}
  \begin{tablenotes}
    \footnotesize
    \item* UIM refers to User Interest Modeling.
    \end{tablenotes}
\end{table}

\subsection{Ablation Study}
To dissect the contribution of each component in CCN, we conduct a thorough ablation study, with results detailed in section (C) of Table \ref{tab:auc}.

Starting from the SIM baseline, adding the trigger feature and standard User Interest Modeling provides a moderate gain of +0.17pt AUC. However, integrating our full CCN framework boosts the performance by an additional +0.33pt, for a total gain of +0.50pt.

Further analysis of CCN's internal components confirms their necessity. Removing the User Interest Modeling, the attraction loss $\mathcal{L}_{\mathcal{S}^{+}}$, or the repulsion loss $\mathcal{L}_{\mathcal{S}^{-}}$ from the full model leads to substantial performance drops, with the AUC gain over baseline decreasing to +0.39pt, +0.36pt, and +0.33pt, respectively.

\subsection{Online A/B Test}
To validate CCN's real-world effectiveness, we conduct online A/B Test during the Taobao Double 11 Shopping Carnival. The results shown in Table \ref{tab:abtest} demonstrate improvements across key metrics. CCN boosts CTR by 12.3\% and order volume by 12.7\%. The concurrent lift in both CTR and orders is noteworthy, as it overcomes the common multi-objective seesaw effect where optimizing for clicks can harm conversions. This indicates that the user preferences learned by CCN are not only accurate for clicks but also translate to further purchase intent. Further online experiments in another scenario shown in Table \ref{tab:abtest_other} confirm that the substantial performance gains are primarily driven by our collaborative contrastive paradigm, far surpassing the benefits of standard feature additions.

\begin{table}[htbp]
  \centering
  \caption{Online A/B test results during the Taobao Double 11 Shopping Carnival. Lifts (\%) are relative to the production baseline (SIM)}
  \label{tab:abtest}
  \small
  \setlength{\tabcolsep}{3pt}
  \begin{tabular}{l c c c c}
    \toprule
    \textbf{Model} & \textbf{CTR} & \textbf{Clicks/User} & \textbf{CVR} & \textbf{Orders/User} \\
    \midrule
    DIHN & -7.3\% & -9.3\% & -1.5\% & -3.1\% \\
    DIAN & -5.6\% & -6.0\% & -0.9\% & -1.5\% \\
    \textbf{CCN (Ours)} & \textbf{+12.3\%} & \textbf{+7.4\%} & \textbf{+0.9\%} & \textbf{+12.7\%} \\
    \bottomrule
  \end{tabular}
\end{table}

\begin{table}[htbp]
  \centering
  \caption{Online A/B test results in another production scenario. Lifts (\%) are relative to the production baseline (SIM).}
  \label{tab:abtest_other}
  \small
  \setlength{\tabcolsep}{4pt}
  \begin{tabular}{l c c c}
    \toprule
    \textbf{Model} & \textbf{CTR} & \textbf{CVR} & \textbf{Visit Purchase Rate} \\
    \midrule
    Trigger + UIM & +2.21\% & +6.37\% & +8.85\%  \\
    \textbf{CCN (Ours)} & \textbf{+17.48\%} & \textbf{+10.06\%} & \textbf{+29.38\%} \\
    \bottomrule
  \end{tabular}
  \begin{tablenotes}
    \footnotesize
    \item* Visit Purchase
Rate here refers to the scenario-guided purchase probability of a daily user view.
    \end{tablenotes}
\end{table}

\section{Conclusion}
In this paper, we address the challenge of recommendation in dynamic and ephemeral trigger-induced scenarios: building robust recommendation models for dynamic and ephemeral trigger-induced scenarios. We identify two fundamental traps in existing methods: trigger-dependency and reliance on long-term data. We propose a new perspective that bypasses ambiguous intent modeling by focusing on direct, in-session collaborative signals. Our proposed framework, the Collaborative Contrastive Network (CCN), operationalizes this perspective. By treating the user-trigger pair as a unique context, CCN effectively learns robust, trigger-conditioned user intent representations from collaborative feedback. Rigorous offline and online experiments on large-scale industrial data validate our approach. In the online A/B test during a previously unseen, large-scale promotional event, CCN achieves a 12.3\% lift in CTR and a 12.7\% lift in order volume over the production baseline. CCN has fully launched across all Taobao's promotional scenarios.

\bibliographystyle{splncs04}
% \balance
\bibliography{article}
\end{document}